**Do PageRank-based author rankings outperform simple citation counts?**

Dalibor Fiala[a, *]

Lovro Šubelj[b]

Slavko Žitnik[b]

Marko Bajec[b]

[a] University of West Bohemia, Department of Computer Science and Engineering

Univerzitní 8, 30614 Plzeň, Czech Republic

[b] University of Ljubljana, Faculty of Computer and Information Science

Večna pot 113, 1000 Ljubljana, Slovenia

\* Corresponding author. Tel.: +420 377 63 24 29.

Email addresses: dalfia@kiv.zcu.cz (D. Fiala), lovro.subelj@fri.uni-lj.si (L. Šubelj), slavko.zitnik@fri.uni-lj.si (S. Žitnik), marko.bajec@fri.uni-lj.si (M. Bajec).

**Abstract**: The basic indicators of a researcher's productivity and impact are still the number of publications and their citation counts. These metrics are clear, straightforward, and easy to obtain. When a ranking of scholars is needed, for instance in grant, award, or promotion procedures, their use is the fastest and cheapest way of prioritizing some scientists over others. However, due to their nature, there is a danger of oversimplifying scientific achievements. Therefore, many other indicators have been proposed including the usage of the PageRank algorithm known for the ranking of webpages and its modifications suited to citation networks. Nevertheless, this recursive method is computationally expensive and even if it has the advantage of favouring prestige over popularity, its application should be well justified, particularly when compared to the standard citation counts. In this study, we analyze three large datasets of computer science papers in the categories of artificial intelligence, software engineering, and theory and methods and apply 12 different ranking methods to the citation networks of authors. We compare the resulting rankings with self-compiled lists of outstanding researchers selected as frequent editorial board members of prestigious journals in the field and conclude that there is no evidence of PageRank-based methods outperforming simple citation counts.

# 1.     Introduction and related work

Ranking researchers has become very popular due to the possible applications in various hiring, promotion, grant, or award procedures, in which manual assessment can be efficiently supplemented with automated techniques. Apart from counting the research money granted, the easiest way to evaluate a researcher's performance is to estimate the quantity and quality of scholarly publications he/she has produced. The former concentrates on production (or productivity) and the latter on impact (or influence). In its basic form, production is the number of research papers a scientist has published and impact is the number of citations from other research publications these papers have attracted. These two simple indicators may already form a basis for an easy ranking of researchers (or authors as all of these evaluations are based on the authorship of research publications). One of the drawbacks of this simplistic approach is that it does not differentiate between popularity and prestige, i.e. it considers all citations as equivalent. In the practice, however, a citation by a Nobel Prize laureate is certainly more valuable than that by a doctoral student, a citation by a scientist with a high number of citations has probably more weight than that by a scholar with only a few citations, and many citations from the same researcher are apparently less worth than the same number of citations from many different scientists. All this motivated the application of "higher-order" evaluation methods (citations being a "first-order" method) such as PageRank to citation networks of authors.

The recursive PageRank algorithm by Brin and Page (1998), the founders of Google, was originally meant to evaluate the importance of webpages on the basis of the link structure of the web. The principal idea is that an important webpage is itself linked to from other important webpages. Thus, a webpage can have a high rank if it has inlinks from many webpages with low ranks but also if it has inlinks from few webpages with high ranks. The rank of a webpage depends on the ranks of the webpages linking to it. In practice, the costly calculation of PageRank in a directed graph is done in an iterative fashion and more on this will be said in the following section. Even though a similar bibliometric concept was introduced by Pinski and Narin (1976) long before Google, the PageRank's property of being applicable to any directed graph was soon utilized in the analysis of citation networks to rank journals (Bollen et al., 2006; Bergstrom, 2007; González-Pereira et al., 2010), papers (Chen et al., 2007; Walker et al., 2007; Ma et al., 2008; Yan and Ding, 2010), authors (Fiala et al., 2008; Ding et al., 2009; Radicchi et al., 2009; Ding, 2011; Fiala, 2011; Yan and Ding, 2011; Fiala, 2012b; Fiala, 2013a; Nykl et al., 2014), a combination of the three (Yan et al., 2011), institutions (Yan, 2014), departments (Fiala, 2013b; Fiala, 2014), countries (Ma et al., 2008; Fiala, 2012a), or a mixture of the above entities (West et al., 2013). In the many previous studies of ours we investigated various PageRank modifications with respect to the standard (baseline) PageRank and concluded that some of the variants performed better than the baseline in that they generated rankings closer to the human perception of a good ranking. In the present study, however, we consider simple citations as the baseline and the main research question is whether author rankings based on PageRank (and its variants) outperform citations in terms of better ranks assigned to outstanding researchers. If the answer was yes, the high computational cost of PageRank needed to overcome some deficiencies of citations would be well justified.

Let us remark in this place that PageRank-based (or, in general, recursive) ranking methods are only one branch of research performance evaluation techniques (in addition to standard publication and citation counts) with the other notable one being the family of h- and g-indices (Hirsch, 2005; Egghe, 2006) that combine both production and impact in a single number. These indices may obviously be used to rank authors as well, but they are not the concern of the present paper which is further organized as follows: In Section 2 we briefly recall the substance of PageRank, its modifications used in our analysis, and other related methods and refer to the relevant literature for more details. In Section 3 we describe the dataset we examined, which consists of papers from three large computer science categories (artificial intelligence, software engineering, and theory & methods). In Section 4 we present and discuss the main results of our analysis and give a negative answer to the main research question asked in the title of this article. And finally, in the last section, we summarize the most important contributions and results of this study and propose some research lines for our future work.

## 2.    Methods

Let us define the directed author citation graph as $G = (V, E)$, where $V$ is the set of vertices (authors) and $E$ is the set of edges (unique citations between authors). If author $v$ cites author $u$ (once or more times), there is an edge $(v, u) \in E$. Then, by the recursive definition, the PageRank score $PR(u)$ of author $u$ depends on the scores of all citing authors in the following way:

$$\mathrm{PR}(u) = \frac{1-d}{|V|} + d \sum_{(v,u) \in E} PR(v)\Omega \qquad (1)$$

where $d$ is the damping factor, which was set to 0.85 in the original web experiments by Brin and Page (1998), and $\Omega$ is either the multiplicative inverse of the out-degree of $v$ like in the standard PageRank or $\sigma_{v,u}\Big/\sum_{(v,k)\in E}\sigma_{v,k}$ like in the bibliographic PageRank by Fiala et al. (2008), where

$$\sigma_{v,k} = w_{v,k}\Big/\Big(\big[(c_{v,k}+1)/(b_{v,k}+1)\big]\sum_{(v,j)\in E}w_{v,j}\Big) \qquad (2)$$

with $w$, $b$, and $c$ being various coefficients determined from both the citation and the collaboration networks of authors which will be explained below. Note that as follows from (1), an author with no citations (incoming edges) will still have a non-zero PageRank, which will be close to the multiplicative inverse of the total number of authors in the dataset. Of course, this will be influenced by the damping factor $d$, which was intitially determined empirically after the observation that a typical web user usually followed five links to other webpages and then chose a random webpage, e.g. by starting a new keyword search, thus resulting in about one sixth ($\approx 0.15$) of all transactions between webpages to be random. Indeed, the total PageRank in the system (or network) should be 1 and the individual PageRanks of vertices are then the fractions of time a random surfer spends there. We refer to the paper by Diligenti et al. (2004) for an explanation of PageRank within a random walk framework. Other approaches to the PageRank problem include solving a linear system (Bianchini et al., 2005; Langville and Meyer, 2004), but for practical reasons it is mostly computed dynamically in an iterative manner until convergence of subsequently generated rankings, which may be measured with Spearman's rank correlation coefficients. This is also the way we applied in our analysis with the maximum number of iterations set to 50, which was enough even with stricter convergence criteria and millions of nodes in the experiment by Brin and Page, and the damping factor set to 0.9 for the calculations to be consistent with our previous studies. (But we also experimented with other damping factors as will be said later in the paper.)

Let us now return to the coefficients $w$, $b$, and $c$ appearing in the bibliographic version of (1) and thus in (2). Their combination will produce a weight of each citation between two authors. The key ideas are the following: a citation between two authors is more intense if it occurs repeatedly ($w_{u,v}$ is the number of all citations from $u$ to $v$); a citation from a colleague (who has coauthored some publications with the cited author) is considered less valuable than a citation from a foreign scientist who has no common papers with the cited author ($c_{u,v}$ is the number of collaborations of $u$ and $v$), and the "collaboration penalty" is mitigated proportionally to some other factors, for instance to the number of coauthors in the joint publications by $u$ and $v$ ($b_{u,v}$ is then the number of common publications by $u$ and $v$). If all the coefficients $b$ and $c$ are set to 0 and $w$ to 1, the bibliographic PageRank becomes the standard PageRank (*PR*) by Brin and Page (1998). If only b's and c's are set to 0, the resulting method is a weighted PageRank (*PR weighted*) similar to that by Xing and Ghorbani (2004). If only $b$'s are set to 0, the variant is called *PR collaboration*. If $b_{u,v}$ is generally non-zero, it can represent one of the following numbers: the number of publications by $u$ plus the number of publications by $v$ (*PR publications*), the number of all coauthors of $u$ plus the number of all coauthors of $v$ (*PR allCoauthors*), the number of all distinct coauthors of $u$ plus the number of all distinct coauthors of $v$ (*PR allDistCoauthors*), the number of publications by $u$ where $u$ is not the only author plus the number of publications by $v$ where $v$ is not the only author (*PR allCollaborations*), the number of coauthors in the common publications by $u$ and $v$ (*PR coauthors*), or the number of distinct coauthors in the common publications by $u$ and $v$ (*PR distCoauthors*). Because it was not the aim of this paper to redefine the bibliographic PageRank and its variants, for its formal definitions we refer to Fiala et al. (2008) and particularly to Fiala (2012).

There is another recursive method related to PageRank which was invented independently of Brin and Page (1998) by Kleinberg (1999). This technique is called HITS and proposes two scores for a webpage, authority and hubness, suggesting that a good authority will be linked to from good hubs and a good hub will link to good authorities. This mutually reinforcing relationship is expressed by the indirect recursion in the following formula:

$$A(u) = \sum_{(v,u) \in E} H(v), \; H(u) = \sum_{(u,v) \in E} A(v) \tag{3}$$

where $A(u)$ is the authority score of u and $H(u)$ its hubness. A close relationship of HITS to PageRank was shown by Ding et al. (2002). We included HITS in our experiments with author rankings and computed iteratively (similarly to PageRank) the authority scores of authors, which were then used to rank them. in descending order. In makes no sense to use the hubness score for the ranking because an author with a high hubness means a highly referencing author, whose prestige, however, may be low.

In addition to the computationally intensive "higher-order" methods PageRank and HITS, we also wanted to rank authors using simple, non-recursive techniques, which are sometimes called "first-order" methods. A prominent representative of this category is the simple citation counting (*Citations*), which is a well established metric of scientific impact and which we will consider as the baseline ranking method. Compared to PageRank, citations are not only cheap in terms of calculation and data collection, but they are also more transparent and easier to understand, which is a big advantage in research assessment. Citations between authors can be easily extracted from the citation networks of papers we had at our disposal. But unlike paper citations that are distinct by nature, there are usually many duplicate citations between authors because researchers often refer to publications on a specific topic covered by a limited set of scholars. So it may well happen that a large number of citations come from a single author. Therefore, it may be useful to count the number of distinct citing authors rather than citations. In the author citation graph (without parallel edges), this number is the in-degree of nodes and we call this method *In-degree* consistently in this study as well as in our earlier articles although alternative names like "CitingAuthors" would also be thinkable.

Thus, in total, we have these 12 author ranking methods: *Citations* (our baseline), *In-degree* (distinct citing authors), *HITS* (authority score), *PR* (standard PageRank), *PR weighted* (weighted PageRank), and bibliographic PageRank variations *PR collaboration*, *PR publications*, *PR allCoauthors*, *PR allDistCoauthors*, *PR allCollaborations*, *PR coauthors*, and *PR allCoauthors* whose rationale is explained above. We will apply these techniques to three large citation networks of computer science authors, generate author rankings, and try to answer the question raised in the title of this article.

## 3.    Data

In middle 2013 we got access to programmatically download XML records with metadata on journal articles and conference papers from the well-known Web of Science (WoS) database. These metadata typically included paper titles, author names, author emails, source titles (journal or conference names), publication years, links to citing papers as well as some other information. We were interested in three subcategories of computer science, namely Artificial Intelligence (AI), Software Engineering (SE), and Theory & Methods (TM), which we wanted to inspect more closely. The choice of these three subcategories was determined by the research interests of the authors of this paper as well as by the necessity to balance the sufficient amount of data for analysis and the time (and costs) needed to acquire these data. Finally, we managed to obtain 179,510 publication records in AI, along with 215,745 records in SE, and 159,107 records in TM. However, these document sets are not disjoint as we can see in Figure 1. This is due to the fact that in the Web of Science papers belong to one or more subject categories or subcategories. Thus, there is an overlap of almost five thousand papers that are classified in each of the three subcategories with a slightly smaller overlap between AI and SE but substantially bigger intersections (by an order of magnitude) between AI and TM on the one hand and SE and TM on the other. In the latter case about a third of the documents are shared by both subcategories. This indicates well that software engineering and theory & methods are two closely related disciplines of computer science. All in all, we analyzed 546,678 publication records in this study.

Insert Figure 1 here.

The publications under investigation span a time period from 1964 to 2013 for AI and from 1954 to 2013 for SE and TM. AI is, therefore, a "younger" discipline than both SE and TM and, of course, year 2013 is incomplete in each case. We can observe in Figure 2 that all disciplines evolved similarly in the course of time and their production gradually increased from a few dozens of papers in the first years to almost 17,000 AI papers in 2006, 9,000 SE papers in 2004, and more than 24,000 TM papers in 2005. (Again, let us recall that the document sets are not disjoint so the total numbers of papers published in the above disciplines are smaller.) We may notice a few remarkable things in Figure 2 and those are the sudden production rise of software engineering publications in the 1980s, the explosion of publication activity in all three areas after 2000 and a rather dramatic general decrease after 2006. This spectacular decline may be partly caused by decreasing governmental budgets due to the approaching global economic and financial crisis but in particular by a change in the indexing strategy of the Web of Science database. This change included, among others, discontinuing the indexation of the well-known "Lecture Notes" book series in the Science Citation Index Expanded.

Insert Figure 2 here.

Before applying ranking methods to authors, we needed to create citation networks of authors from the citation networks of papers. Between the papers in AI there were 639,126 citations, in SE there were 323,444 citations, and in TM there were 483,603 citations. We extracted publications' authors and linked together the authors of each citing and cited paper removing self-citations. In this way, we obtained 119,430 authors linked by 4,349,759 citations in AI, 108,079 authors with 2,118,037 citations in SE, and 123,656 authors with 3,248,792 citations in TM. Let us note in this place that no name unification or disambiguation was performed, which would have been extremely time-consuming regarding the big volume of data we analyzed. The authors were only identified by their full surnames and first name and middle name initials, which was the usual way they were supplied in our WoS data. All in all, our primary goal was to present general ranking features rather than individual ranks although these are also provided for the reader's reference in the appendix.

Comparing author rankings is always tricky as there are no "ground truth values" for the ranks that would tell us whether a ranking method works well or not. The only viable option if such a standard (or reference) ranking does not exist is to have a reference set of "good" authors about whom we know that they should be ranked high by a good ranking and low by a "bad" ranking. We may compile a list of outstanding authors based on the winners of some prestigious computer science awards (Sidiropoulos and Manolopoulos, 2005; Fiala et al., 2008; Fiala, 2011; Fiala, 2012) or on the editorial board members of some prestigious computer science journals (a similar concept employed by Liu et al., 2005), which we have done in this study because there are no compatible awards in artificial intelligence, software engineering, and theory & methods. To this end, we manually inspected the editorial boards of the top ten journals by impact factor in the 2012 edition of Journal Citation Reports® (Thomson Reuters, 2013) in the three aforementioned categories. After some minimum data cleaning we included in our reference set of significant authors in each area those who appeared on more than one editorial board and checked these names for ambiguities. At the end of this process, we obtained 32, 12, and 17 authors whose names can be seen in Tables A.4, A.5, and A.6 in the appendix.

## 4.     Results and discussion

We applied all of the twelve ranking methods mentioned in Section 3 to the author citation networks in AI, SE, and TM and obtained 12 different author rankings. The ranking methods are *Citations*, *In-degree*, *HITS*, (standard) PageRank (*PR*), weighted PageRank (*PR weighted*) and seven other PageRank variants described earlier. Figure 3 depicts boxplots of author rankings in each category showing the relative ranks (to be able to compare networks of different sizes) achieved by the best, worst, and median editorial board member from the reference set of outstanding researchers in a discipline. Relative ranks are calculated by dividing the original ranks by the number of authors in each network (AI, SE, and TM) to always fall between 0 and 1. This is a very simple way how rankings with different numbers of authors may be compared. Alternatively, a ranking function might be used for the comparison of these rankings such as the normalized discounted cumulative gain (Järvelin and Kekäläinen, 2002), but its more costly computation would likely not result in a better visualization than the boxplots in Figure 3. As is usual with boxplots, the top edge of each bar marks the 75th percentile of the ranks assigned to the outstanding scholars by a particular ranking method and the bottom edge of each bar represents the 25th percentile. The short line dividing each box into two sections is the median rank. Please note that the lower the rank the better the position of a researcher because, obviously, rank 1 is better than rank 100 when speaking in absolute terms. (An optimum ranking, if there is one, would place all the authors from the reference set to top positions, e.g. 1 – 32 in AI, and the box in its boxplot would be virtually invisible in Figure 3.) There is also a straight line in each section of the chart denoting the median rank yielded by the simple citation counting, which we consider a baseline. As we may notice, PageRank-based variants always have a worse median rank than citations except for *PR allCoauthors* and *PR allDistCoauthors* in TM, where, however, they still have much worse maximum ranks. These two variants take into account the number of all (distinct) coauthors in the common publications of the citing and cited author and perform comparably well (but not better than) citations in SE. However, their reputation as the best PageRank variants does not hold in AI in which they perform worse than the other PageRank-like methods. Thus, it is inconclusive and we cannot say which PageRank-based methods are the best, but we can almost certainly claim that, on the basis of our experiment, there is no evidence that author ranking methods similar to PageRank outperform simple (and much cheaper) citation counts.

Insert Figure 3 here.

What is somewhat striking is the poor performance of HITS in SE but actually quite good scores in AI and TM. So, again, it is unclear whether HITS is better or worse than citations based on this experiment, similarly to our previous studies (Fiala et al. 2008; Fiala, 2011; Fiala, 2012). On the other hand, the good performance of *In-degree* seems to be quite stable in Figure 3 where it slightly outperforms *Citations* in all three citation networks. (Let us recall that in *In-degree* citations from one author are counted only once so a high rank in *In-degree* may better indicate how well-known an author is in the community than simple citations. This feature of *In-degree* seems to be crucial for editorial board members.) To get some additional support for these conclusions, we ran another set of experiments the main results of which may be seen in Figure 4. Despite the lack of compatible awards in the three disciplines under study and a different evaluation methodology of choice for the present analysis we mentioned earlier, this time the reference set of researchers, whose ranks yielded by various ranking methods we compared, consisted of 28 ACM A.M. Turing Award winners from 1991 to 2010 as described in Fiala (2012). As we may note, the median ranks achieved in AI are quite high and very low in SE and particularly in TM, which indicates that the Turing Award is more relevant for the latter two categories. Indeed, even the worst positions of the awardees based on TM data are still in the better half of the rankings, in contrast to AI and SE. And in addition, while there is no award winner missing in the TM rankings, there is one omission in SE and even 15 laureates missing in AI. Thus, although the PageRank-related methods perform roughly the same as simple citations in AI and TM and somewhat better in SE, due to the missing data and unequal relevance of the three computer science categories for the selected assessment methodology, we may probably conclude again that there is no evidence that PageRank-based rankings would outperform citation counts.

Insert Figure 4 here.

Let us note at this place that we carried out the whole analysis with the damping factor set to 0.9 for the study to be compatible with our earlier research, but we also tested a damping factor set to 0.5 as proposed by Chen et al. (2007), Walker et al. (2007), Ma et al. (2008) or Ding et al. (2009) to find out that even if performing slightly better, PageRank variants are still far from outperforming simple citations. The exact ranks along with aggregate values underlying Figure 3 are shown in Tables A.1, A.2, and A.3 in the appendix. The values of the baseline method (*Citations*) are typeset in italics and the aggregate values that are better than baseline are highlighted in bold. In the other tables in the appendix (Tables A.4, A.5, and A.6), we show the top 30 researchers in AI, SE, and TM as calculated by *Citations*, *In-degree*, *HITS*, (standard) PageRank (*PR*), and the most different PageRank variant (*PR allCoauthors*). HITS and PageRank scores are also presented (although they depend on many factors like the convergence criterion, damping factor, etc.) for the reader to get a clue how wide or narrow the gaps between the ranks are. But we will not discuss the standings of the individual authors in detail because the aim of this analysis was to evaluate various ranking methods as a whole rather than to assess individuals. As for the PageRank variant whose ranks differ most from the standard PageRank (*PR allCoauthors*), we found it by comparing pairwise Spearman correlations of the 12 rankings in each of the three computer science categories. From the heatmaps in Figure 5 it is quite obvious that there are three groups of rankings: *Citations* and *In-degree* are, as expected, very closely related as are PageRank and its modifications while HITS is a stand-alone category. However, even though all the correlations are very high (more than 0.8), we must be aware that this is true for rankings with well over 100,000 authors. Rankings with far fewer authors (e.g. 100, 500, or 1000), which are much more common in reality, would very probably have considerably lower correlations.

Insert Figure 5 here.

Let us now return to the evaluation methodology again. Besides editorial board members (or conference programme committee members, which is the same in essence but less appropriate with WoS data where conference papers are known to be absent or scarcely present) as the reference set of outstanding researchers, an alternative approach are lists of various computer science award winners. As we have said, we used this methodology successfully in the past (Fiala et al., 2008; Fiala, 2011; Fiala, 2012) and although the research goals set in those studies were different, it is easy to check that even then the PageRank-based methods mostly did not perform better than simple citations. In this analysis, we intentionally avoid prize awardees in order to test the viability of the current approach with editorial board members. Regarding author name disambiguation, even though no merging and/or unmerging of author names was performed prior to the analysis and the WoS data were treated "as is", we believe that the results of our study are still valid. We have shown in our earlier work (Fiala, 2011) that analyzing even much more inconsistent CiteSeer data may lead to relevant conclusions. And while we recognize that some of the names presented in the tables in the appendix may need disambiguation or merging (as may also some others in lower positions not shown there), their individual ranks are actually not so important as the aggregate values displayed in Figure 3. As none of the ranking methods applied disambiguates author names, we expect the overall trend not to change even if all of them did.

Finally, let us speculate a little bit about the reasons of the disappointing performance of the PageRank-based methods as compared to simple citations. The most straightforward explanation seems to be that the evaluation methodology (editorial board membership) itself relies on pure citations. This appears to be a valid point since members of journal editorial boards are usually persons of high repute, well known in their scientific community, who publish frequently and are often cited by other researchers. The same is certainly true also for conference programme committee chairs or members or for computer science award winners. On the other hand, PageRank and related techniques are concerned with the quantity of citations as well as with their quality. They reflect prestige rather than popularity. In this context, it would seem that the editorial board members of the journals we selected for our analysis were chosen on the basis of popularity rather than prestige. Interestingly, a similar observation may be made for the award winners in our previous studies (e.g. Fiala et al, 2008), where, however, the baseline method was the standard PageRank and not citations. Even in studies where author credit was distributed in a slightly different (West et al., 2013) or a more different (Radicchi et al., 2009) way, a high correlation with simple citations was reported. We can see no reason why this bias of the assessment methodology towards simple citations should be absent when conference programme committee members are used as a reference set. In fact, all thinkable evaluation approaches (including peer judgement) are based on citations to some extent and we are not aware of any exception. If such an exceptional approach existed, it would be interesting to run our experiments again and see if the outcome is different.

## 5.    Conclusions and future work

The quality of researchers is often assessed using basic scientometric indicators like the number of publications and citations and even though many other more advanced metrics have been proposed, in principle they always rely on the publication output and impact of a scholar. One of these more advanced techniques is the PageRank algorithm which was originally conceived to rank webpages but has been successfully used to evaluate authors of research papers as well. This algorithm is recursive in nature and requires dozens of iterations over the whole citation network to generate a stable ranking of authors. Thus, it is quite costly compared to simply counting citations and the key question is whether it is worth of it. Does PageRank benefit author rankings compared to citation counts? In this study we tried to address this problem and our response to the question is negative. In particular, we made the following contributions:

- We created large citation networks of authors from 179,510 papers in artificial intelligence, 215,745 papers in software engineering, and 246,391 papers in theory & methods  - subfields of computer science - by programmatically querying the Web of Science database.
- We compiled lists of editorial board members of prestigious journals in each category to have three reference sets of outstanding researchers and generated 12 rankings of authors using various methods including citation counts, PageRank, and its modifications.
- We compared the rankings with each other by visualizing their basic statistics on boxplot charts and depicting their correlations on heatmaps.

The main findings of our study are the following:

- There is no evidence of PageRank-based author rankings outperforming simple citation counts in terms of better mean or median ranks assigned to the authors in a reference set of prestigious scholars in a computer science category.
- From the PageRank modifications, the variant with considering all coauthors in common publications of the citing and cited authors seems to work best. The performance of HITS is unstable and the ranking that takes into account citations only from distinct authors (*In-degree*) appears to yield better results than standard citation counts.
- All PageRank-based rankings are very highly correlated with each other, while HITS and citations-based rankings are the other two distinct ranking groups. Still, all the 12 rankings under study are rather strongly correlated with Spearman's rho being 0.8 at least.

In our future work, we would like to concentrate also on other categories of computer science or on other scientific fields. We intend to extend our experiments and further investigate some phenomena we observed in this analysis such as the circumstances in which *In-degree* performs better or worse than citations or HITS performs better or worse than PageRank. In addition to editorial board members, who may themselves be selected based on their citation counts, another set of experiments should be run with different reference sets of outstanding authors, e.g. with researchers receiving a prestigious award in a particular domain of computer science or another research area. Another line of research may include investigations whether simple citations outperform also some other well established evaluation metrics such as the h-index.

**Acknowledgements**

Thanks are due to Thomson Reuters for providing us with the data. For D. Fiala, this work was supported by the European Regional Development Fund (ERDF), project "NTIS - New Technologies for Information Society", European Centre of Excellence, CZ.1.05/1.1.00/02.0090 and in part by the Ministry of Education of the Czech Republic under grant MSMT MOBILITY 7AMB14SK090. For L. Šubelj, S. Žitnik, and M. Bajec, this work was supported in part by the Slovenian Research Agency Program No. P2-0359, by the Slovenian Ministry of Education, Science and Sport Grant No. 430-168/2013/91, and by the European Union, European Social Fund.

**Figure captions**

**Fig. 1**     Venn diagram showing the numbers of documents in artificial intelligence (AI), software engineering (SE), and theory & methods (TM) categories

**Fig. 2**     Numbers of publications in artificial intelligence (AI), software engineering (SE), and theory & methods (TM) categories in individual years

**Fig. 3**     Boxplots depicting relative ranks achieved by various ranking methods for artificial intelligence (left), software engineering (centre), and theory & methods (right) editorial board members, with the horizontal lines marking the median rank yielded by the "baseline" method (simple citation counts) in each category

**Fig. 4**     Boxplots depicting relative ranks achieved by various ranking methods for artificial intelligence (left), software engineering (centre), and theory & methods (right) ACM A. M. Turing Award winners, with the horizontal lines marking the median rank yielded by the "baseline" method (simple citation counts) in each category

**Fig. 5**     Heatmaps of pairwise Spearman correlations of all rankings in artificial intelligence (AI), software engineering (SE), and theory & methods (TM) categories

**Table captions**

**Table A.1**  Top artificial intelligence editorial board members and their ranks achieved by various ranking methods

**Table A.2**  Top software engineering editorial board members and their ranks achieved by various ranking methods

**Table A.3**  Top theory & methods editorial board members and their ranks achieved by various ranking methods

**Table A.4**  Top 30 artificial intelligence researchers by citations, in-degree, HITS, PageRank and the most different PageRank variant

**Table A.5**  Top 30 software  engineering researchers by citations, in-degree, HITS, PageRank and the most different PageRank variant

**Table A.6**  Top 30 theory & methods researchers by citations, in-degree, HITS, PageRank and the most different PageRank variant

**Appendix A**

Insert Table A.1 here.

Insert Table A.2 here.

Insert Table A.3 here.

Insert Table A.4 here.

Insert Table A.5 here.

Insert Table A.6 here.

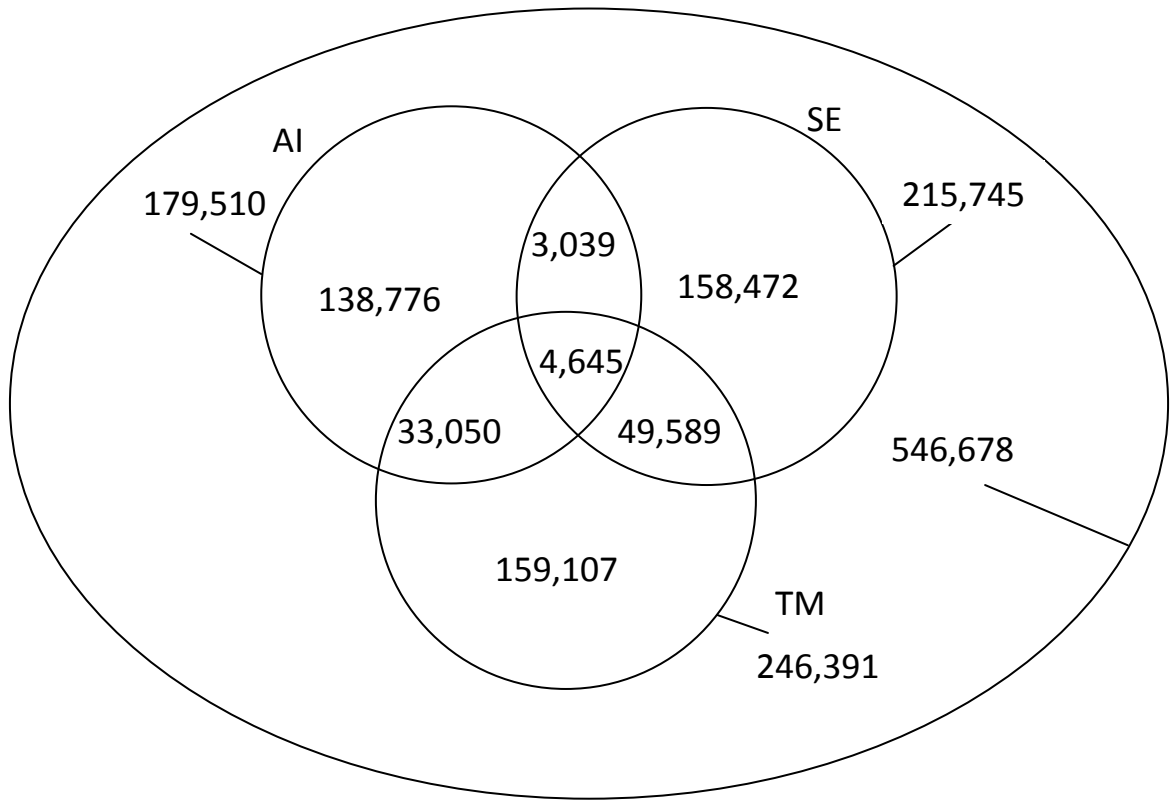

**Figure 2**

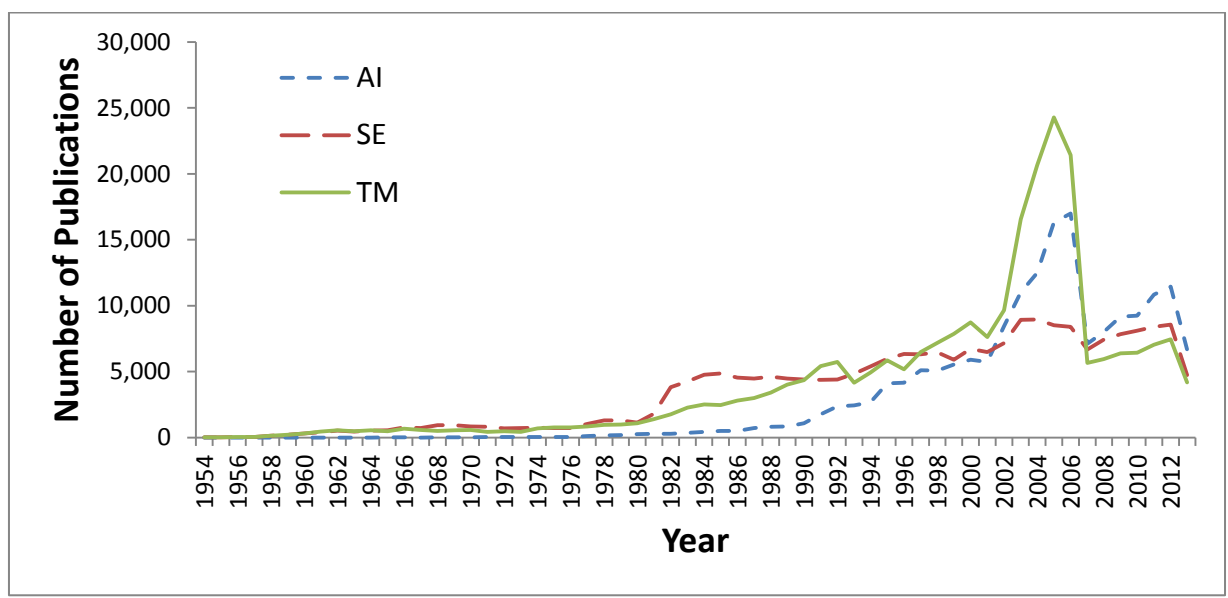

**Figure 3**

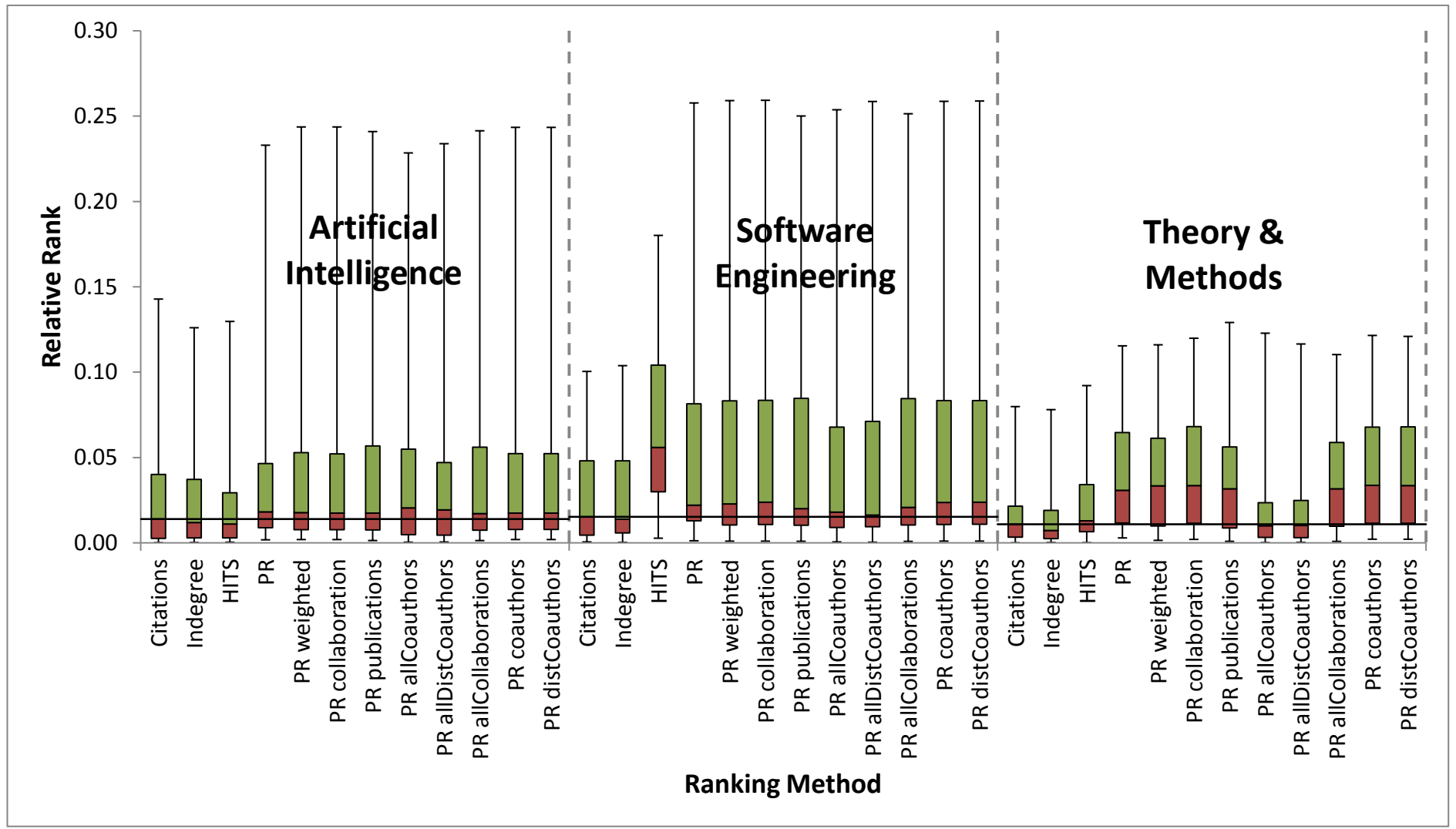

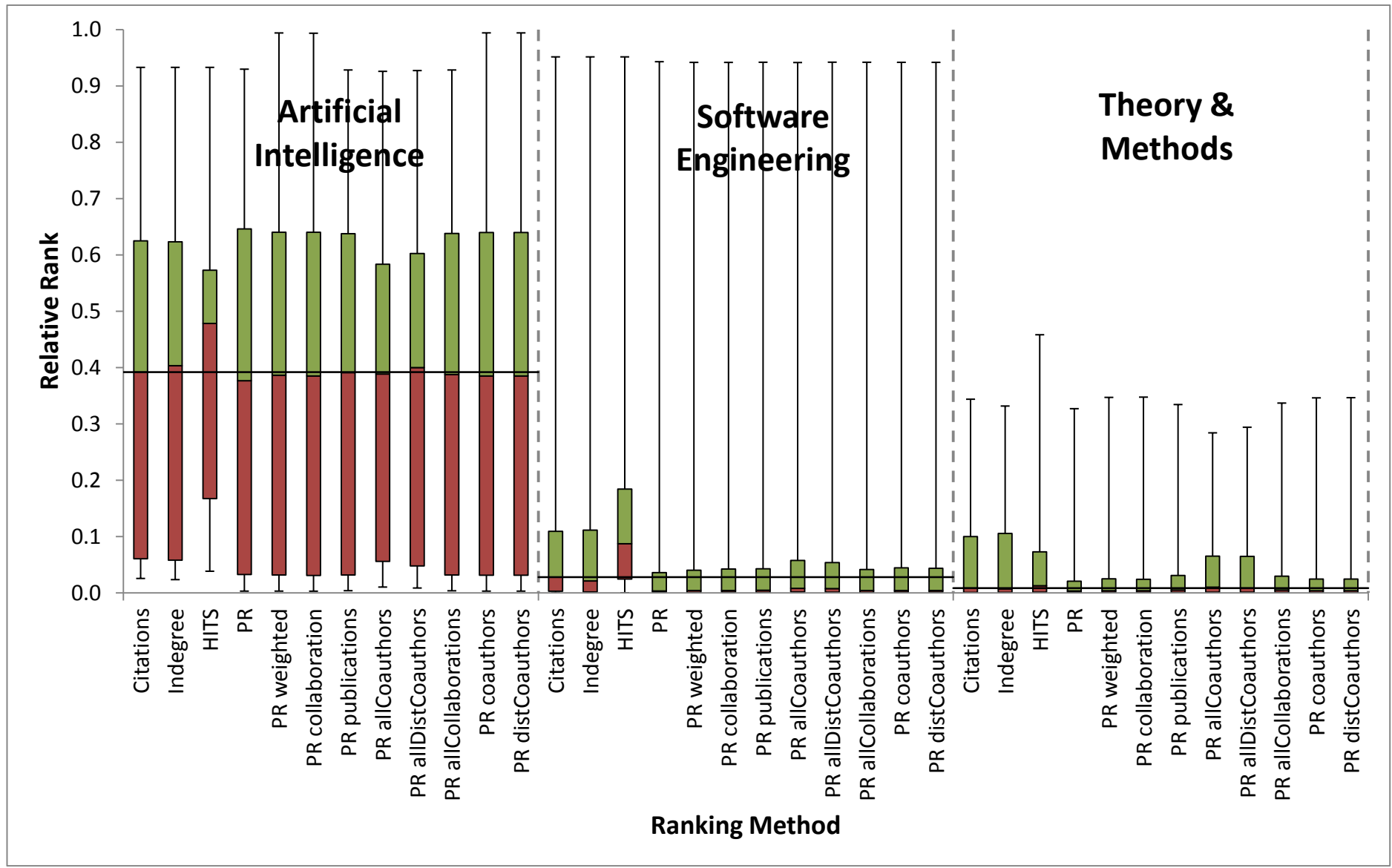

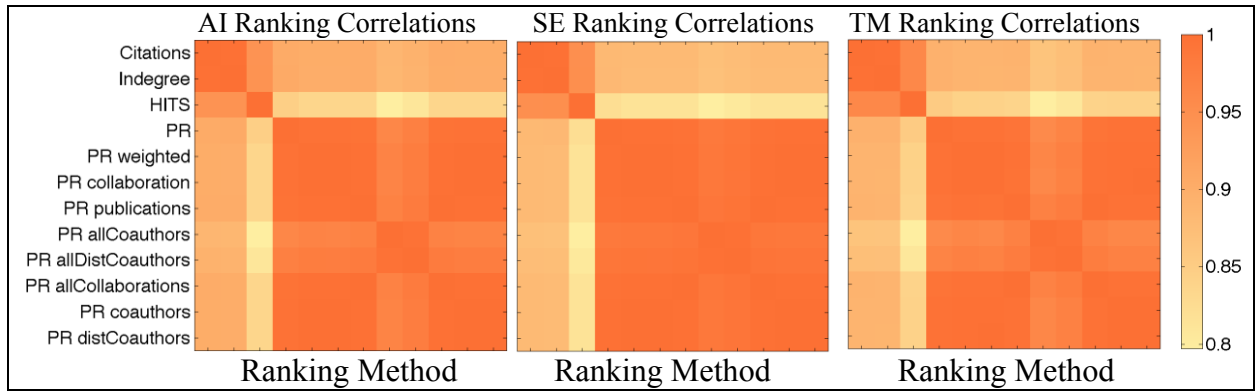

**Fig. 5**     Heatmaps of pairwise Spearman correlations of all rankings in artificial intelligence (AI), software engineering (SE), and theory & methods (TM) categories

**Table A.1**     Top artificial intelligence editorial board members and their ranks achieved by various ranking methods

| Author | Citat-ions | In-degree | HITS | PR | PR weight-ed | PR collab-oration | PR public-ations | PR allCo-authors | PR allDistCo-authors | PR allCollab-orations | PR co-authors | PR distCo-authors |
|---|---|---|---|---|---|---|---|---|---|---|---|---|
| Abbass, H | *2,511* | 2,791 | 3,130 | 8,424 | 8,337 | 8,603 | 8,022 | 2,256 | 2,664 | 8,199 | 8,505 | 8,503 |
| Bach, F | *1,248* | 1,083 | 944 | 1,311 | 1,395 | 1,349 | 1,465 | 3,627 | 3,380 | 1,447 | 1,361 | 1,361 |
| Bregler, C | *5,621* | 6,430 | 4,523 | 7,333 | 6,474 | 6,321 | 6,826 | 15,633 | 13,735 | 6,747 | 6,255 | 6,265 |
| Brown, M | *2,238* | 1,847 | 1,827 | 2,986 | 3,110 | 3,041 | 3,260 | 4,356 | 4,261 | 3,235 | 3,041 | 3,038 |
| Collins, R | *1,021* | 836 | 516 | 1,562 | 1,807 | 1,777 | 1,632 | 2,742 | 2,459 | 1,631 | 1,789 | 1,792 |
| Cordon, O | *519* | 732 | 1,115 | 3,487 | 2,816 | 2,887 | 2,905 | 1,921 | 2,041 | 2,892 | 2,866 | 2,863 |
| Herrera, F | *20* | 64 | 221 | 840 | 518 | 564 | 543 | 115 | 192 | 535 | 567 | 567 |
| Ishibuchi, H | *196* | 311 | 470 | 1,130 | 931 | 947 | 919 | 440 | 548 | 931 | 948 | 947 |
| Ishikawa, H | *2,612* | 3,006 | 2,133 | 1,900 | 1,306 | 1,280 | 1,438 | 2,645 | 2,310 | 1,415 | 1,288 | 1,286 |
| Kim, JH | *240* | 202 | 182 | 601 | 707 | 721 | 480 | 220 | 203 | 484 | 726 | 727 |
| Learned-Miller, E | *10,287* | 8,965 | 7,059 | 7,650 | 8,709 | 8,622 | 8,606 | 14,287 | 12,928 | 8,535 | 8,607 | 8,604 |
| Li, X | *108* | 237 | 156 | 1,464 | 1,116 | 1,219 | 735 | 106 | 109 | 723 | 1,221 | 1,220 |
| Liu, D | *968* | 995 | 1,144 | 3,106 | 3,236 | 3,197 | 3,322 | 1,987 | 2,177 | 3,313 | 3,227 | 3,211 |
| Lu, J | *283* | 463 | 192 | 935 | 770 | 731 | 855 | 383 | 395 | 853 | 760 | 757 |
| Matsushita, Y | *9,374* | 8,723 | 4,015 | 10,601 | 11,167 | 11,337 | 11,636 | 6,494 | 6,188 | 11,519 | 11,389 | 11,359 |
| Mori, G | *3,402* | 2,991 | 1,817 | 5,669 | 6,775 | 7,002 | 7,282 | 3,005 | 3,738 | 7,246 | 7,007 | 7,005 |
| Navab, N | *4,639* | 4,310 | 4,746 | 5,511 | 6,252 | 6,200 | 6,760 | 7,563 | 7,071 | 6,685 | 6,242 | 6,238 |
| Ong, YS | *257* | 341 | 591 | 2,040 | 1,596 | 1,767 | 1,618 | 95 | 88 | 1,639 | 1,776 | 1,776 |
| Pal, NR | *47* | 36 | 54 | 221 | 239 | 238 | 171 | 56 | 70 | 168 | 240 | 241 |
| Panella, M | *13,653* | 12,638 | 9,662 | 27,826 | 29,104 | 29,103 | 28,780 | 27,286 | 27,932 | 28,835 | 29,075 | 29,074 |
| Pedrycz, W | *122* | 110 | 208 | 619 | 622 | 651 | 327 | 169 | 174 | 405 | 661 | 660 |
| Pennec, X | *1,369* | 1,183 | 1,401 | 1,724 | 1,723 | 1,739 | 1,864 | 1,485 | 1,452 | 1,835 | 1,742 | 1,741 |
| Ramanan, D | *1,972* | 1,675 | 1,224 | 3,012 | 2,854 | 2,859 | 2,286 | 2,320 | 2,297 | 2,252 | 2,850 | 2,853 |
| Roth, S | *3,161* | 2,528 | 3,333 | 2,298 | 2,715 | 2,660 | 2,792 | 5,153 | 4,542 | 2,768 | 2,656 | 2,658 |
| Sato, Y | *2,668* | 2,025 | 2,126 | 1,062 | 1,126 | 1,105 | 1,167 | 2,584 | 2,239 | 1,171 | 1,114 | 1,110 |
| Skrjanc, I | *7,506* | 7,767 | 15,498 | 13,279 | 13,090 | 13,299 | 12,829 | 7,857 | 9,121 | 12,901 | 13,294 | 13,298 |
| Sutton, C | *17,064* | 15,058 | 12,722 | 5,237 | 5,803 | 5,871 | 5,643 | 7,369 | 6,747 | 5,645 | 5,736 | 5,768 |
| Torralba, A | *453* | 448 | 383 | 1,038 | 896 | 874 | 914 | 2,127 | 1,866 | 894 | 880 | 878 |
| Vemuri, BC | *329* | 283 | 244 | 310 | 287 | 280 | 336 | 631 | 563 | 328 | 283 | 283 |
| Welling, M | *5,245* | 4,816 | 2,901 | 2,826 | 2,420 | 2,374 | 2,551 | 6,710 | 5,441 | 2,520 | 2,380 | 2,373 |
| Williams, C | *363* | 343 | 269 | 234 | 247 | 249 | 278 | 621 | 493 | 270 | 248 | 249 |
| Zhao, D | *7,181* | 6,665 | 4,885 | 13,043 | 11,287 | 11,214 | 11,546 | 16,970 | 16,078 | 11,470 | 11,243 | 11,233 |
| **mean rank** | *3,334* | 3,122 | 2,803 | 4,352 | 4,357 | 4,378 | 4,368 | 4,663 | 4,484 | 4,359 | 4,374 | 4,373 |
| **median rank** | *1,671* | 1,429 | 1,313 | 2,169 | 2,114 | 2,076 | 2,075 | 2,452 | 2,304 | 2,044 | 2,085 | 2,083 |
| **min. rank** | *20* | 36 | 54 | 221 | 239 | 238 | 171 | 56 | 70 | 168 | 240 | 241 |
| **max. rank** | *17,064* | 15,058 | 15,498 | 27,826 | 29,104 | 29,103 | 28,780 | 27,286 | 27,932 | 28,835 | 29,075 | 29,074 |
| **std. deviation** | *4,188* | 3,821 | 3,666 | 5,523 | 5,692 | 5,709 | 5,685 | 6,023 | 5,879 | 5,689 | 5,702 | 5,701 |

**Table A.2**    Top software engineering editorial board members and their ranks achieved by various ranking methods

| Author | Citat-ions | In-degree | HITS | *PR* | PR weight-ed | PR collab-oration | PR public-ations | PR allCo-authors | PR allDistCo-authors | PR allCollab-orations | PR co-authors | PR distCo-authors |
|---|---|---|---|---|---|---|---|---|---|---|---|---|
| Bertino, E | *839* | 652 | 3,463 | 2,941 | 3,033 | 3,161 | 2,707 | 1,631 | 1,703 | 2,848 | 3,147 | 3,144 |
| Blake, MB | *8,130* | 9,806 | 19,472 | 26,215 | 26,903 | 26,841 | 26,756 | 27,425 | 27,949 | 26,833 | 26,836 | 26,843 |
| Boneh, D | *10,861* | 11,226 | 16,476 | 27,857 | 28,005 | 28,024 | 27,032 | 23,581 | 24,605 | 27,173 | 27,959 | 27,979 |
| Clarke, S | *2,707* | 2,928 | 8,537 | 7,008 | 6,478 | 6,909 | 4,265 | 1,681 | 1,679 | 4,864 | 6,899 | 6,920 |
| Dustdar, S | *2,390* | 2,028 | 5,830 | 6,552 | 7,117 | 7,172 | 7,372 | 3,911 | 4,373 | 7,344 | 7,181 | 7,176 |
| Forsyth, D | *387* | 532 | 303 | 1,446 | 1,028 | 1,012 | 1,092 | 2,211 | 1,817 | 1,104 | 1,041 | 1,030 |
| Ghezzi, C | *413* | 246 | 2,566 | 822 | 989 | 1,020 | 651 | 447 | 508 | 654 | 984 | 999 |
| Gottlob, G | *896* | 944 | 6,262 | 1,817 | 1,883 | 1,975 | 1,614 | 1,004 | 1,067 | 1,642 | 1,978 | 2,006 |
| Jouppi, N | *9,322* | 8,419 | 18,865 | 14,183 | 14,603 | 14,554 | 14,462 | 17,584 | 17,641 | 14,494 | 14,519 | 14,505 |
| Morrisett, G | *511* | 663 | 4,462 | 1,557 | 1,170 | 1,209 | 1,117 | 900 | 866 | 1,136 | 1,200 | 1,228 |
| Wing, J | *68* | 38 | 2,096 | 140 | 116 | 116 | 106 | 70 | 49 | 101 | 117 | 118 |
| Wright, MH | *4,209* | 4,122 | 9,503 | 1,285 | 1,463 | 1,429 | 1,600 | 3,116 | 2,632 | 1,535 | 1,419 | 1,406 |
| **mean rank** | *3,394* | 3,467 | 8,153 | 7,652 | 7,732 | 7,785 | 7,398 | 6,963 | 7,074 | 7,477 | 7,773 | 7,780 |
| **median rank** | *1,643* | 1,486 | 6,046 | 2,379 | 2,458 | 2,568 | 2,161 | 1,946 | 1,760 | 2,245 | 2,563 | 2,575 |
| **min. rank** | *68* | 38 | 303 | 140 | 116 | 116 | 106 | 70 | **49** | 101 | 117 | 118 |
| **max. rank** | *10,861* | 11,226 | 19,472 | 27,857 | 28,005 | 28,024 | 27,032 | 27,425 | 27,949 | 27,173 | 27,959 | 27,979 |
| **std. deviation** | *3,714* | 3,879 | 6,379 | 9,454 | 9,638 | 9,613 | 9,517 | 9,456 | 9,723 | 9,536 | 9,601 | 9,602 |

**Table A.3**    Top theory & methods editorial board members and their ranks achieved by various ranking methods

| Author | Citat-ions | In-degree | HITS | *PR* | PR weight-ed | PR collab-oration | PR public-ations | PR allCo-authors | PR allDistCo-authors | PR allCollab-orations | PR co-authors | PR distCo-authors |
|---|---|---|---|---|---|---|---|---|---|---|---|---|
| Liu, Y | *417* | 309 | 964 | 1,778 | 1,747 | 1,806 | 1,453 | 333 | 321 | 1,478 | 1,795 | 1,794 |
| Wing, J | *453* | 340 | 528 | 699 | 573 | 564 | 679 | 603 | 571 | 613 | 579 | 572 |
| Gottlob, G | *233* | 230 | 309 | 990 | 1,024 | 1,079 | 653 | 405 | 381 | 681 | 1,099 | 1,099 |
| Morrisett, G | *9,879* | 9,659 | 11,399 | 14,279 | 14,353 | 14,829 | 15,970 | 15,195 | 14,413 | 13,646 | 15,033 | 14,964 |
| Boneh, D | *14* | 37 | 21 | 366 | 189 | 259 | 118 | 59 | 49 | 114 | 273 | 272 |
| Crowcroft, J | *5,990* | 4,926 | 3,878 | 12,312 | 13,402 | 13,355 | 15,719 | 6,507 | 7,118 | 13,382 | 13,566 | 13,503 |
| Beyer, HG | *145* | 155 | 872 | 1,432 | 838 | 1,229 | 1,086 | 29 | 57 | 1,195 | 1,239 | 1,244 |
| Dorigo, M | *524* | 496 | 1,606 | 2,434 | 2,127 | 2,302 | 2,123 | 775 | 727 | 2,122 | 2,292 | 2,301 |
| Lozano, JA | *1,920* | 1,641 | 4,696 | 7,560 | 7,031 | 8,423 | 6,850 | 1,247 | 1,251 | 7,267 | 8,389 | 8,398 |
| Miller, J | *192* | 143 | 722 | 1,454 | 1,230 | 1,417 | 988 | 177 | 253 | 1,196 | 1,418 | 1,417 |
| Suganthan, PN | *2,777* | 2,735 | 6,428 | 9,520 | 8,861 | 9,204 | 7,388 | 2,905 | 3,065 | 7,555 | 9,102 | 9,119 |
| Tan, KC | *2,657* | 2,352 | 4,412 | 5,019 | 5,525 | 5,455 | 5,706 | 3,166 | 3,841 | 5,776 | 5,482 | 5,478 |
| Zhang, M | *7,037* | 5,734 | 2,612 | 12,081 | 12,999 | 12,887 | 12,308 | 7,953 | 8,277 | 12,383 | 12,814 | 12,819 |
| Zhang, J | *1,354* | 908 | 2,521 | 3,804 | 4,137 | 4,149 | 3,916 | 1,427 | 1,481 | 3,913 | 4,166 | 4,157 |
| Li, X | *1,950* | 1,531 | 1,150 | 5,128 | 5,880 | 5,870 | 5,557 | 2,299 | 2,676 | 5,627 | 5,949 | 5,926 |
| Ong, YS | *1,561* | 1,395 | 4,212 | 7,993 | 7,569 | 7,754 | 6,954 | 1,979 | 1,910 | 7,083 | 7,684 | 7,704 |
| Wu, J | *572* | 421 | 802 | 1,385 | 1,546 | 1,528 | 1,382 | 652 | 608 | 1,390 | 1,548 | 1,546 |
| **mean rank** | *2,216* | 1,942 | 2,772 | 5,190 | 5,237 | 5,418 | 5,226 | 2,689 | 2,765 | 5,025 | 5,437 | 5,430 |
| **median rank** | *1,354* | 908 | 1,606 | 3,804 | 4,137 | 4,149 | 3,916 | **1,247** | **1,251** | 3,913 | 4,166 | 4,157 |
| **min. rank** | *14* | 37 | 21 | 366 | 189 | 259 | 118 | 29 | 49 | 114 | 273 | 272 |
| **max. rank** | *9,879* | 9,659 | 11,399 | 14,279 | 14,353 | 14,829 | 15,970 | 15,195 | 14,413 | 13,646 | 15,033 | 14,964 |
| **std. deviation** | *2,735* | 2,519 | 2,820 | 4,458 | 4,661 | 4,718 | 5,026 | 3,808 | 3,736 | 4,486 | 4,745 | 4,733 |

**Table A.4**      Top 30 artificial intelligence researchers by citations, in-degree, HITS, PageRank and the most different PageRank variant

| | Citations | | In-degree | | HITS [*10⁻²] | | PR [*10⁻⁴] | | PR allCoauthors [*10⁻³] | |
|---|---|---|---|---|---|---|---|---|---|---|
| 1 | Jain, AK | 15,021 | Jain, AK | 6,810 | Jain, AK | 13.1856 | Horn, BKP | 26.7431 | Jain, AK | 18.9662 |
| 2 | Malik, J | 8,607 | Malik, J | 4,417 | Malik, J | 10.2278 | Ballard, DH | 26.6847 | Kittler, J | 12.9174 |
| 3 | Kittler, J | 8,238 | Kittler, J | 4,344 | Kittler, J | 9.8324 | Geman, S | 25.9486 | Wang, L | 8.3717 |
| 4 | Kriegman, DJ | 7,541 | Scholkopf, B | 3,600 | Kriegman, DJ | 9.1965 | Hornik, K | 23.4745 | Wang, J | 8.1870 |
| 5 | Scholkopf, B | 7,332 | Vapnik, V | 3,593 | Scholkopf, B | 8.9430 | Geman, D | 22.9776 | Kanade, T | 7.1650 |
| 6 | Kanade, T | 7,032 | Kanade, T | 3,586 | Belhumeur, PN | 8.7021 | Jain, AK | 21.5837 | Oja, E | 6.2053 |
| 7 | Duin, RPW | 6,670 | Lowe, DG | 3,520 | Duin, RPW | 8.5493 | White, H | 20.6053 | Huang, TS | 6.1994 |
| 8 | Belhumeur, PN | 6,471 | Duin, RPW | 3,330 | Kanade, T | 8.5259 | Fikes, RE | 20.5439 | Scholkopf, B | 5.8157 |
| 9 | Vapnik, V | 6,075 | Breiman, L | 3,147 | Lowe, DG | 8.0279 | Nilsson, NJ | 19.8517 | Zhang, D | 5.0933 |
| 10 | Breiman, L | 6,006 | Jordan, MI | 3,069 | Muller, KR | 7.9694 | Funahashi, K | 19.1373 | Duin, RPW | 4.6992 |
| 11 | Muller, Kr | 5,729 | Kriegman, DJ | 3,046 | Vapnik, V | 7.6957 | Haralick, RM | 18.5182 | Jain, A | 3.9943 |
| 12 | Lowe, DG | 5,725 | Hornik, K | 3,043 | Poggio, T | 7.3748 | Stinchcombe, M | 18.4966 | Poggio, T | 3.7520 |
| 13 | Geman, D | 5,028 | Geman, S | 2,978 | Matas, J | 7.1384 | Brooks, RA | 16.7612 | Nayar, SK | 3.6473 |
| 14 | Lin, CJ | 5,019 | Muller, KR | 2,932 | Hespanha, JP | 7.0629 | Pearl, J | 16.4261 | Rosenfeld, A | 3.2826 |
| 15 | Jordan, MI | 4,987 | Lin, CJ | 2,918 | Meer, P | 6.8320 | Pentland, AP | 15.7140 | Bigun, J | 3.0795 |
| 16 | Sejnowski, TJ | 4,921 | Geman, D | 2,863 | Huang, TS | 6.7858 | Terzopoulos, D | 14.7983 | Bischof, H | 3.0754 |
| 17 | Bezdek, JC | 4,914 | Belhumeur, PN | 2,723 | Jordan, MI | 6.7426 | Rosenfeld, A | 14.0835 | Hornik, K | 2.9340 |
| 18 | Hornik, K | 4,749 | Wang, L | 2,679 | Breiman, L | 6.3566 | Vapnik, V | 13.4566 | Plaza, E | 2.5533 |
| 19 | Zhang, D | 4,732 | Sejnowski, TJ | 2,651 | Taylor, CJ | 6.3527 | Schunck, BG | 13.4545 | Weickert, J | 2.4774 |
| 20 | Herrera, F | 4,724 | Poggio, T | 2,604 | Lin, CJ | 6.3022 | Canny, J | 13.4450 | Leonardis, A | 2.4613 |
| 21 | Geman, S | 4,711 | Ballard, DH | 2,579 | Geman, D | 6.2773 | Newell, A | 13.3067 | Sycara, K | 2.3444 |
| 22 | Grossberg, S | 4,630 | Rosenfeld, A | 2,562 | Geman, S | 6.2021 | Malik, J | 13.1673 | Rangarajan, A | 2.1807 |
| 23 | Poggio, T | 4,626 | Meer, P | 2,539 | Zhang, D | 6.1346 | Grossberg, S | 13.1471 | Amari, S | 2.1802 |
| 24 | Schmid, C | 4,585 | Huang, TS | 2,533 | Wang, L | 6.0553 | Davis, R | 13.0938 | Matas, J | 2.1711 |
| 25 | Taylor, Cj | 4,475 | Canny, J | 2,498 | Cootes, TF | 5.9839 | Lowe, DG | 13.0680 | Kimmel, R | 2.1366 |
| 26 | Rosenfeld, A | 4,427 | Cortes, C | 2,494 | Canny, J | 5.9161 | Sejnowski, TJ | 13.0536 | Muller, KR | 2.0960 |
| 27 | Oja, E | 4,332 | Matas, J | 2,479 | Pentland, AP | 5.8954 | Hinton, GE | 13.0220 | Dorigo, M | 2.0939 |
| 28 | Huang, TS | 4,312 | Haralick, RM | 2,472 | Ballard, DH | 5.8614 | Huang, TS | 12.6856 | Liu, J | 2.0708 |
| 29 | Jennings, NR | 4,199 | Pentland, AP | 2,415 | Cortes, C | 5.8490 | Davis, LS | 12.5888 | Zisserman, A | 2.0354 |
| 30 | Horn, BKP | 4,181 | Oja, E | 2,381 | Sejnowski, TJ | 5.8004 | | | Bro, R | 2.0325 |

**Table A.5**   Top 30 software engineering researchers by citations, in-degree, HITS, PageRank and the most different PageRank variant

| | Citations | | In-degree | | HITS [*10⁻²] | | PR [*10⁻⁴] | | PR allCoauthors [*10⁻³] | |
|---|---|---|---|---|---|---|---|---|---|---|
| 1 | Basili, VR | 4,155 | Lamport, L | 2,372 | Cohen-Or, D | 14.4513 | Hoare, CAR | 49.7986 | Hoare, CAR | 14.0238 |
| 2 | Lamport, L | 4,022 | Hoare, CAR | 2,213 | Seidel, HP | 13.1369 | McCarthy, J | 34.9504 | Wirth, N | 11.5275 |
| 3 | Cohen-Or, D | 3,903 | Basili, VR | 2,002 | Shum, HY | 12.8951 | Kammerer, HC | 31.8550 | Seidel, HP | 9.4998 |
| 4 | Hoare, CAR | 3,385 | Harel, D | 1,814 | Guo, BN | 11.1822 | Oktay, S | 31.8550 | Cohen-Or, D | 4.3613 |
| 5 | Seidel, HP | 3,075 | Shamir, A | 1,730 | Turk, G | 10.9546 | Dennis, JB | 29.6339 | Shum, HY | 4.0523 |
| 6 | Shamir, A | 3,025 | Seidel, HP | 1,663 | Desbrun, M | 10.5037 | Codd, EF | 27.8712 | Basili, VR | 3.8873 |
| 7 | Shum, HY | 2,942 | Cohen-Or, D | 1,653 | Zhou, K | 9.4780 | Dijkstra, EW | 27.5035 | Harel, D | 3.8712 |
| 8 | Harel, D | 2,749 | Parnas, DL | 1,543 | Hoppe, H | 9.3236 | Wirth, N | 24.9930 | Parnas, DL | 3.3924 |
| 9 | Briand, LC | 2,559 | Shum, HY | 1,375 | Gross, M | 9.3158 | Denning, PJ | 24.7114 | Denning, PJ | 3.2165 |
| 10 | Guo, BN | 2,333 | Tarjan, RE | 1,354 | Alexa, M | 9.2896 | Parnas, DL | 23.3554 | McCarthy, J | 3.1794 |
| 11 | Kemerer, CF | 2,285 | Kemerer, CF | 1,274 | Rusinkiewicz, S | 9.2514 | Floyd, RW | 23.2359 | Vardi, MY | 3.1321 |
| 12 | Parnas, DL | 2,233 | Turk, G | 1,192 | Durand, F | 9.1238 | Perlis, AJ | 21.8722 | Dennis, JB | 3.0947 |
| 13 | Weyuker, EJ | 2,229 | Weiser, M | 1,108 | Sheffer, A | 8.6284 | Chu, RC | 21.4381 | Dijkstra, EW | 3.0317 |
| 14 | Tarjan, RE | 2,183 | Weyuker, EJ | 1,104 | Bao, HJ | 8.4892 | Hwang, UP | 21.4381 | Emerson, EA | 2.7773 |
| 15 | Desbrun, M | 2,097 | Dijkstra, EW | 1,103 | Sorkine, O | 8.4684 | Simons, RE | 21.4381 | Shamir, A | 2.5844 |
| 16 | Turk, G | 2,036 | Briand, LC | 1,101 | Kobbelt, L | 8.4185 | Lamport, L | 20.1893 | Flanagan, C | 2.4894 |
| 17 | Gross, M | 1,928 | Guo, BN | 1,087 | Hu, SM | 8.3370 | Knuth, DE | 17.0970 | Clarke, EM | 2.4707 |
| 18 | Fedkiw, R | 1,814 | Desbrun, M | 1,083 | Pauly, M | 8.2149 | Horst, R | 16.7542 | Ball, T | 2.3321 |
| 19 | Cignoni, P | 1,776 | Cignoni, P | 1,055 | Shamir, A | 8.2064 | Hoffman, KL | 16.6144 | Tarjan, RE | 2.3015 |
| 20 | Reps, T | 1,756 | Reps, T | 1,051 | Cignoni, P | 8.1103 | Mizell, AM | 15.6718 | Alur, R | 2.2184 |
| 21 | Levoy, M | 1,747 | Levoy, M | 1,049 | Lischinski, D | 7.9935 | Tarjan, RE | 15.6258 | Guo, BN | 2.1979 |
| 22 | Zhou, K | 1,728 | Clarke, EM | 1,006 | Curless, B | 7.9242 | Weiss, RA | 15.5402 | Gross, M | 2.1701 |
| 23 | Durand, F | 1,695 | Gross, M | 999 | Gotsman, C | 7.6949 | Phong, BT | 15.4064 | Durand, F | 2.1210 |
| 24 | Weiser, M | 1,665 | Durand, F | 979 | Levy, B | 7.4184 | Shamir, A | 14.4281 | Lamport, L | 2.1072 |
| 25 | Alexa, M | 1,646 | Scopigno, R | 979 | Alliez, P | 7.3908 | Gries, D | 14.3008 | Montanari, U | 2.0659 |
| 26 | Kobbelt, L | 1,621 | Lee, J | 971 | Ju, T | 7.3030 | Harel, D | 13.0491 | Floyd, RW | 2.0252 |
| 27 | Rusinkiewicz, S | 1,577 | Ferrante, J | 918 | Yu, YZ | 7.3017 | Landin, PJ | 12.3079 | Todd, SJP | 1.9993 |
| 28 | Scopigno, R | 1,569 | Kramer, J | 894 | Popovic, J | 7.2730 | Metcalfe, RM | 12.0450 | Chamberlin, DD | 1.8682 |
| 29 | Clarke, EM | 1,564 | Alexa, M | 890 | Warren, J | 7.2027 | Robinson, JA | 11.8506 | Codd, EF | 1.8481 |
| 30 | Hoppe, H | 1,541 | Hoppe, H | 881 | Scopigno, R | 7.1808 | Boggs, DR | 11.5930 | Lee, J | 1.7246 |

**Table A.6**    Top 30 theory & methods researchers by citations, in-degree, HITS, PageRank and the most different PageRank variant

| | Citations | | In-degree | | HITS [*10⁻²] | | PR [*10⁻⁴] | | PR allCoauthors [*10⁻³] | |
|---|---|---|---|---|---|---|---|---|---|---|
| 1 | Tarjan, RE | 9,586 | Tarjan, RE | 3,587 | Tarjan, RE | 10.9182 | Hoare, CAR | 37.8686 | Yung, M | 14.5603 |
| 2 | Shamir, A | 8,108 | Lamport, L | 3,315 | Shamir, A | 10.8676 | Perlis, AJ | 32.1174 | Alur, R | 7.4745 |
| 3 | Lamport, L | 7,324 | Shamir, A | 3,035 | Rivest, RL | 10.4166 | Codd, EF | 30.4506 | Vardi, MY | 6.2202 |
| 4 | Micali, S | 6,653 | Zadeh, LA | 2,810 | Yannakakis, M | 9.6018 | Dennis, JB | 30.1616 | Dongarra, J | 6.0710 |
| 5 | Alur, R | 6,389 | Rivest, RL | 2,513 | Yung, M | 9.3902 | Tarjan, RE | 29.2852 | Shamir, A | 6.0548 |
| 6 | Yannakakis, M | 5,429 | Papadimitriou, CH | 2,375 | Papadimitriou, CH | 9.2467 | Dijkstra, EW | 27.6138 | Hoare, CAR | 6.0394 |
| 7 | Rivest, RL | 5,354 | Valiant, LG | 2,357 | Lamport, L | 8.9736 | McCarthy, J | 24.0778 | Preneel, B | 5.7847 |
| 8 | Goldwasser, S | 5,270 | Hoare, CAR | 2,351 | Micali, S | 8.7176 | Shamir, A | 22.4969 | Deb, K | 5.6748 |
| 9 | Valiant, LG | 5,265 | Yannakakis, M | 2,351 | Goldwasser, S | 8.5713 | Knuth, DE | 21.9525 | Pnueli, A | 5.2145 |
| 10 | Papadimitriou, CH | 5,104 | Foster, I | 2,138 | Goldreich, O | 8.4746 | Zadeh, LA | 21.7615 | Dongarra, JJ | 4.9583 |
| 11 | Bellare, M | 4,981 | Alur, R | 1,926 | Bellare, M | 8.2476 | Denning, PJ | 21.3309 | Tarjan, RE | 4.7165 |
| 12 | Yung, M | 4,943 | Yung, M | 1,900 | Valiant, LG | 7.9701 | Lamport, L | 21.1109 | Nielsen, M | 4.5767 |
| 13 | Milner, R | 4,905 | Fischer, MJ | 1,899 | Naor, M | 7.4863 | Walden, DC | 20.6401 | Micali, S | 3.7363 |
| 14 | Boneh, D | 4,585 | Milner, R | 1,855 | Stern, J | 7.3887 | Wirth, N | 20.6227 | Rozenberg, G | 3.5906 |
| 15 | Zadeh, LA | 4,371 | Floyd, S | 1,704 | Luby, M | 7.3704 | Rivest, RL | 20.5698 | Rivest, RL | 3.5245 |
| 16 | Goldreich, O | 4,339 | Vardi, MY | 1,692 | Fischer, MJ | 7.0054 | Floyd, RW | 19.3385 | Zadeh, LA | 3.2584 |
| 17 | Stern, J | 4,286 | Parrow, J | 1,660 | Parrow, J | 6.9366 | Valiant, LG | 17.2807 | Naor, M | 3.1258 |
| 18 | Henzinger, TA | 4,280 | Kesselman, C | 1,657 | Alur, R | 6.6362 | Adleman, L | 15.5315 | Bellare, M | 3.0279 |
| 19 | Hoare, CAR | 4,167 | Dongarra, JJ | 1,602 | Preneel, B | 6.6117 | Vanhorn, EC | 15.3883 | Kumar, V | 2.8850 |
| 20 | Vardi, MY | 4,099 | Micali, S | 1,557 | Krawczyk, H | 6.5033 | Robinson, JA | 15.3753 | Henzinger, TA | 2.8087 |
| 21 | Deb, K | 4,024 | Dongarra, J | 1,553 | Boneh, D | 6.5033 | Parnas, DL | 13.9488 | Grumberg, O | 2.7709 |
| 22 | Foster, I | 3,833 | Bellare, M | 1,539 | Fiat, A | 6.3567 | Milner, R | 13.6916 | Hennessy, M | 2.7249 |
| 23 | Krawczyk, H | 3,725 | Tuecke, S | 1,531 | Adleman, L | 6.3541 | Fischer, MJ | 13.1500 | Bergstra, JA | 2.6503 |
| 24 | Camenisch, J | 3,556 | Adleman, L | 1,530 | Johnson, DS | 6.3340 | Ullman, JD | 12.9868 | Krawczyk, H | 2.6414 |
| 25 | Parrow, J | 3,533 | Goldreich, O | 1,506 | Okamoto, T | 6.2513 | Gries, D | 12.7681 | Goldreich, O | 2.5898 |
| 26 | Fischer, MJ | 3,532 | Goldwasser, S | 1,502 | Abadi, M | 6.2247 | Papadimitriou, CH | 12.3012 | Stern, J | 2.5846 |
| 27 | Kaliski, BS | 3,455 | Johnson, DS | 1,492 | Wigderson, A | 6.2152 | Blum, M | 12.0470 | Colchester, A | 2.5771 |
| 28 | Preneel, B | 3,376 | Pnueli, A | 1,481 | Franklin, M | 6.1176 | Yannakakis, M | 11.8855 | Dubois, D | 2.4958 |
| 29 | Dill, DL | 3,368 | Jain, AK | 1,473 | Maurer, U | 6.1106 | Gustafson, RN | 11.7996 | Beyer, HG | 2.4750 |
| 30 | Okamoto, T | 3,224 | Ullman, JD | 1,463 | Vardi, MY | 6.0960 | Sparacio, FJ | 11.7996 | Yager, RR | 2.4538 |